# Multiple Subset Problem as an encryption scheme for communication

Yair Zadok[1], Nadav Voloch[2], Noa Voloch-Bloch[3] and Maor Meir Hajaj[4]

[1] Achva Academic College, MP. Shikmim, Israel
[2] Ruppin Academic Center, Emek Hefer, Israel
[3] Raicol Crystals Ltd., Rosh Ha'Ayin, Israel
[4] University of Haifa, Israel
`voloch@post.bgu.ac.il`

**Abstract.** Using well-known mathematical problems for encryption is a widely used technique because they are computationally hard and provide security against potential attacks on the encryption method. The subset sum problem (SSP) can be defined as finding a subset of integers from a given set, whose sum is equal to a specified integer. The classic SSP has various variants, one of which is the multiple-subset problem (MSSP). In the MSSP, the goal is to select items from a given set and distribute them among multiple bins, ensuring that the capacity of each bin is not exceeded while maximizing the total weight of the selected items. This approach addresses a related problem with a different perspective. Here a related different kind of problem is approached: given a set of sets $A = \{A_1, A_2,...,A_n\}$, find an integer $s$, for which every subset of the given sets is summed up to, if such an integer exists. The problem is NP-complete when considering it as a variant of SSP. However, there exists an algorithm that is relatively efficient for known private keys. This algorithm is based on dispensing non-relevant values of the potential sums. In this paper we present the encryption scheme based on MSSP and present its novel usage and implementation in communication.

## 1   Introduction and related work

Methods of encryption have been widely used in the past decades for various purposes, from financial transactions, working with databases, internet communication, etc. Different encryption schemes fit different purposes. The basic encryption idea is to create a scheme that will be secure from possible untrustworthy third parties, that intend to approach our data. For that, the data being transferred is usually in a form that is not understandable to someone that does not have the knowledge of the encryption algorithm, or certain keys, that the receiver of the message does have. In this paper we further develop our encryption scheme that is based on a variant of the Multiple Subset Problem and was presented in [1]. We implement this method as a secure communication protocol, that has a unique and novel method of encrypting and decrypting data.

The subset sum problem (SSP) is a specific instance of the knapsack problem within the field of computational complexity theory. It encompasses various variations and is known to be NP-complete (as described in [2], that analyzes all sorts of NP-complete problems). The classic problem, as shown in [3], is described as follows: Let $A = \{a_1, a_2, ..., a_m\}$, $a_i \in \mathbb{N}$, and $s \in \mathbb{N}$, find $B \subseteq A$, for which $\sum_{a_i \in B} a_i = s$. Meaning for a given a set of integers $A$, and an integer $s$, find a subset of items from $A$ summing up to $s$.

SSP, being *NP-complete,* has a naïve exponential time solution of cycling all of the possible subsets, and a slightly better ($O(2^{n/2})$) solution shown in research that handles Computing Partitions with Applications to the Knapsack Problem [4].

The existence of subsets that add up to the required sum can be determined using pseudo-polynomial dynamic programming solutions, which utilize an array with Boolean values, resulting in more efficient solutions, as presented in [5], that shows a new technique called *balancing* for the solution of Knapsack Problems. The research proves that an optimal solution to the Knapsack Problem is balanced, and thus only balanced feasible solutions need to be enumerated in order to solve the problem to optimality. [6] presents a new algorithm that is the fastest general algorithm for this problem, and a modified algorithm for cyclic groups, which computes all the realizable subset sums within the group. Efficient approximations algorithms are also shown in [7], that covers different approximation algorithms for combinatorial problems, Subset Sum is one of them, and [8], that focuses on the Subset Sum problem, and find a complete polynomial approximation scheme for it, and shows their results accordingly. An alternative form of SSP is the multiple-subset problem (MSSP), wherein a set of items is chosen and distributed among multiple identical bins, ensuring that the capacity of each bin is not surpassed, while aiming to maximize the total weight of the items. Approximation algorithms for MSSP were demonstrated in [9], which is the first research that formulates the problem and its possible approximation solutions. A special case of MSSP is the equal subset problem (ESS), in which given a set $A = \{a_1, a_2, ..., a_m\}$, $a_i \in \mathbb{N}$, and $s \in \mathbb{N}$, the problem is to decide whether two disjoint subsets $A_1, A_2 \subseteq A$ exist whose elements sum up to the same value, meaning $\sum_{a_i \in A_1} a_i = s$ and $\sum_{a_i \in A_2} a_i = s$ . In [10], the authors introduce a problem and propose an approximation algorithm to solve it. [11] discusses various problems and variations derived from ESS, while [12] presents an efficient approximation algorithm for it.

There is some novel state-of-the-art encryption research such as [13], that describes the principles and methods underlying the creation of an application in secure operating systems, which provides reliable data encryption. The research aims to analyze and indicate the specifics of encryption methods and algorithms based on domestic standards in open-source operating systems. Cryptanalysis was used in the article, as this avoids vulnerabilities identified in previously created implementations. In the article, the authors draw attention to the fact that 7-Zip uses CBC encryption (concatenation of encrypted text blocks), but the Counter Mode is supported. The same support was provided in the encrypt implementation. Since the key expansion function initially fills the special array created by p7zip with round keys using a unique property of the domestic standard, only one round encryption function was created (performed both during encryption and decryption).

This method is also used in various modes. In many cases, initialization time deviations depending on the selected mode are insignificant. The created cryptographic module was tested to meet the domestic standard, which contains several test cases. It was confirmed during the tests that the created module implements the algorithm of the domestic standard. The article shows a way to implement a fairly convenient graphical interface for accessing the cryptographic module, which enables the user not to call the command line and remember the sequence and types of parameters passed to p7zip. This implementation also takes into account the verification of the correctness of decryption and the reading of other error codes.

There are also different types of encryptions, like a hybrid encryption, that is describes and used in another state-of-the-art scheme in [14], that discusses the integration of Internet of Things (IoT) with cloud computing (CC) to improve efficiency in service delivery. The integration is achieved through the development of an integrated IoT system with cloud computing, using a hybrid encryption mechanism for security. The proposed system has been implemented and its performance has been evaluated using various metrics such as power consumption, packet delivery ratio, and algorithm execution time. The system also shows resilience against attacks like the black hole attack.

In this paper, we focus on a different problem that is similar to ESS but deals with a collection of sets instead of a single set of integers. Additionally, we aim to determine an unknown parameter (s) for the sum. This problem has a cryptographic application, which will be discussed in the subsequent sections of this paper.

While previous studies have addressed ESS and provided efficient approximation algorithms for it, there is a need to explore and propose solutions for this related problem and its cryptographic implications. It is this research gap that the current study seeks to fill.

Our contribution of this paper is of several aspects: Firstly, we formulate the Multiple Integrated Subset Problem (MISSP), with examples of it. Second, we describe an encryption methodology scheme based on this problem, and last, we show a communication implementation that uses this encryption scheme.

## 2  Multiple Integrated Subset problem

A problem arises when ESS is addressed in a different manner that is as follows: For a given family of sets of integers $A = \{A_1, A_2, ..., A_n\}$, find an integer $s$, such that for every set $A_i$ some of its subsets is summed up to $s$, if such an integer exists. We denote this problem as the *MISSP*- Multiple Integrated Subset Problem.
The similarity between this problem and ESS lies in the requirement of finding a comparable sum in various subsets. However, the difference lies in the constraints where the problem deals with a collection of sets (2D array) instead, implying a predefined and consistent number of elements in each part of the problem (each set).

Furthermore, the sum ($s$) is a parameter that is not known and needs to be determined for each set. This condition leads to an $O(n)$ complexity increase as SSP needs to be performed for every possible value of s.

The subsequent sections of this paper will demonstrate an enhancement in efficiency for this increase. Table 1 shows a few examples of integrated sets on a small scale. As seen, implementing SSP for finding sums on multiple integrated sets is not injective, thus multiple results of *s* can appear (like in set *C*), or none (like in set *B*).

The probability of obtaining at least one result of *s* decreases as the number of sets (*n*) increases, since there are fewer opportunities for the sets to have equal sums. Conversely, the probability of obtaining at least one result of s increases as the number of items in each set (*m*) increases, as there are more possibilities for equal sums between the sets. This is because there are inevitably more options of creating such a sum. This problem has a cryptographic application when the sets have equal sizes (*m*) and each integer in the sets has the same number of digits (*d*). In this application, the cipher text is represented by the sets, the private keys consist of two of the parameters *m, n,* and *d,* and the encrypted plain text is the resulting sum s. In order to accurately decrypt the cipher text, it is necessary to obtain a single result of *s*.

The complete algorithm for efficiently solving MISSP, along with its proof of correctness and analysis of its complexity, can be found in our previous work ([1]).

## 3    Encryption with MISSP

The classic SSP in cryptography, known as the Merkle-Hellman knapsack cryptosystem ([15]), was one of the earliest created. It was derived from the knapsack problem with SSP as a special case ([16]). This fundamental cryptosystem relied on constructing a super-increasing set, where each number is larger than the sum of all preceding numbers. The SSP was then solved using a greedy polynomial time algorithm ([17]). However, this cryptosystem was later compromised, leading to the development of more intricate versions of SSP for stronger cryptosystems ([18]). In this context, we propose the utilization of MISSP as a relatively robust cryptosystem based on two symmetric private keys.

**Table 1.** Small scale examples for MISSP with unknown parameters

| 2D set | Set | Set items | *s* | summations |
|---|---|---|---|---|
| *A* | $A_1$ | {22, 4, 23, 16} | 49 | 22+23+4=49 |
|  | $A_2$ | {8, 3, 17, 21} | 49 | 17+21+8+3=49 |
|  | $A_3$ | {8, 13, 9, 19} | 49 | 13+9+8+19=49 |
| *B* | $B_1$ | {22, 3, 20, 15} | None | None |
|  | $B_2$ | {5, 1, 17, 21} | None | None |
|  | $B_3$ | {8, 10, 7, 19} | None | None |
|  | $B_4$ | {23, 5, 26, 19, 4} | None | None |
| *C* | $C_1$ | {8, 15, 11, 9, 1} | 10; 21 | 9+1=10; 9+1+11=21 |
|  | $C_2$ | {13, 2, 7, 1} | 10; 21 | 7+2+1=10; 7+1+13=21 |
|  | $C_3$ | {18, 11, 10,19} | 10; 21 | 10=10; 10+11=21 |

*MISSP* has a cryptographic application for equal sizes of sets (*m*), from which we derive the number of sets-*n*, and an equal number of digits of every integer in the sets (*d*). In this particular application, the cipher text refers to the series of provided sets, while the private symmetric keys consist of two parameters: *m, n,* and *d*.

The encrypted plain text is represented by the sum s obtained from these parameters. To ensure an accurate decryption of the cipher text, a singular result of *s* is essential. Fig.1 demonstrates the decryption process, which comprises three distinct steps. The first step involves decomposing the cipher text into n distinct sets. In the example of the figure, the cipher text is 55495458205016966826278532461565, $K_n$ (that is the first private key, representing *n* in the algorithm, that is the number of sets) - is 4, hence the cipher text is decomposed to {55495458, 20501696, 68262785, 32461565}. The second stage is decomposing the resulted sets to *d*-size items. In the example $K_d$=2 ($K_d$ is the second private key, representing *d* in the algorithm, that is the number of digits of every integer in the sets) hence the sets are decomposed to the 2D set of 2-digit integers, on which the *MISSP* algorithm will run, and it is {{55, 49, 54, 58}, {20, 50, 16, 96}, {68, 26, 27, 85}, {32, 46, 15, 65}}. For the last part we run *MISSP* on the 2D set. Its result is the actual plain text. In the example it is 112 for the following summations: 54+58, 96+16, 85+27, and 65+15+32.

There are two important points to note.

Firstly, the selection of keys *n* and *d* is essentially the same as any combination of *m* and *d* or *m*. Secondly, the decomposition process yields the exact same 2D set, albeit with different stages of decomposition. Additionally, the division of the cipher text (*C*) can be conducted either for the integers or for the sets first. This is attributed to the fact that the size of the cipher text, denoted as *|C|,* is equal to *mnd*, where m represents the item size, *n* represents the size of each set, and *d* represents the number of sets.

Second, there is singularity constraint for the resulted plain text, meaning *MISSP* necessarily has to have exactly one result. For example, set *A* in table 1 is suitable for *MISSP* encryption (1 result), but sets *B* (0 results) and *C* (2 results) are not suitable.

We can see the reverse process of decryption of the same values (112 as plain text to 55495458205016966826278532461565 as the cypher text in Fig.2, using, of course, the same private keys for the different parameters of *MISSP*.

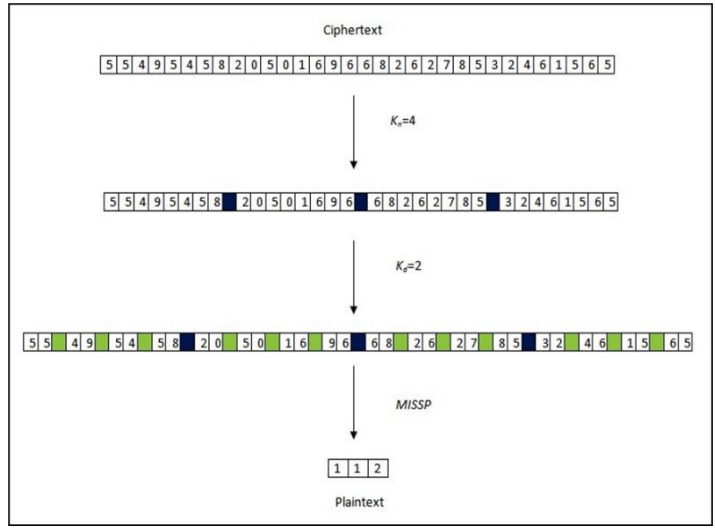

**Fig. 1.** *MISSP* cipher text decomposed to sets of integers by the private keys and decrypted by the *MISSP* algorithm.

## 4      Results

Today, there are accepted and well-known methods for encrypting over the Internet. These methods are known and are a target for hacking and attacks by many factors, such as private hackers, organizations and even countries, so it is important to renew these methods all the time.

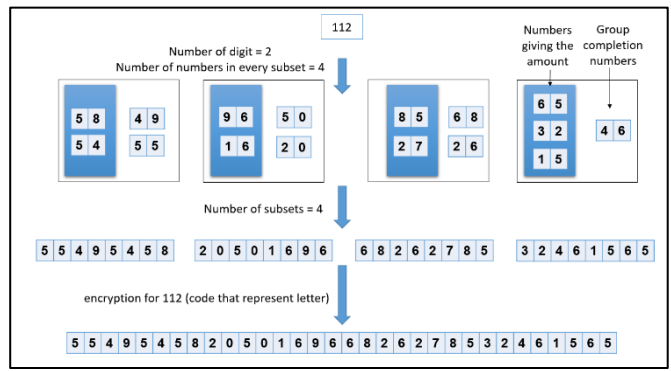

**Fig. 2.**  Plain text encrypted to sets of integers by the private keys by the *MISSP* algorithm.

For the purpose of this research, we have developed a communication application that incorporates a modern form of encryption that is not based on an existing mechanism. The application performs the encryption and decryption based on the *MISSP* algorithm. The app is installed on the Client's end point, and it enables loading the relevant files (dictionary, keys, and data file), and it and then selecting an encryption or decryption operation.

Once the desired operation is completed, it is possible to send the file created through the network.

The basic technological architecture of the communication application is described in Fig.3. As we can see in the figure, for the UI and algorithm implementation .NET technology was used, and for the communication protocol we used Oracle's Virtual Box.

In Fig.4 we can see the class application design for the communication protocol, with the following classes:

Sign Class - for defining a sign in the dictionary. original - the original sign codeNumber - code in the dictionary.

Dictionary Class - for defining a dictionary that the insert text (that we want to encrypt) is written in. dictionary - a list of Signs.

Encryption Class - The class contains the encryption algorithm, contines the encryption keys and dictionary, returns encrypted text.

Decryption Class - The class contains the decryption algorithm, contines the decryption keys and dictionary, returns decrypted text.

We achieved a fully operational application that displays all relevant data, sends and receives data, and accepts the user preferences. The code for this application is presented in [19]. In the implementation itself, we have devised a dictionary for the plaintext to be established, creating some known equal subsets for the purpose of encryption. In table 2 and 3 we can see more results and examples of using MSSIP encryption.

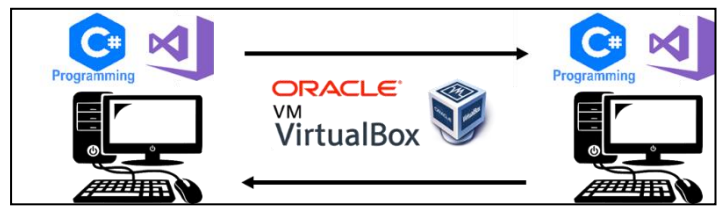

**Fig. 3.** Communication application system architecture for encryption with *MISSP* algorithm.

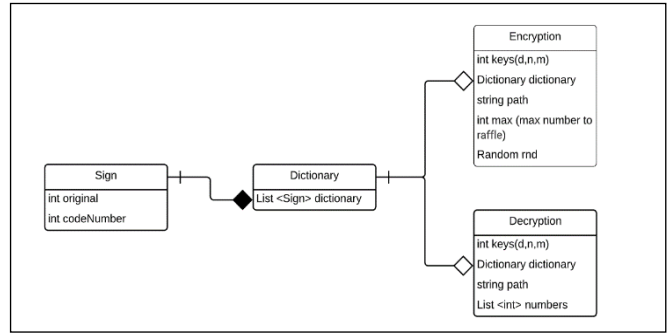

**Fig. 4.** Class Application design for *MISSP* algorithm communication encryption.

**Table 2.** MISSP decryption for C=
35493131748813154458736528552740704862298228292954554958985451406684422196443311513849292218285679345148168531612583692956543931393165971925901260799466515318973136998975301895

| n/m | d | sets | Set items | s (plain text) | summations |
|---|---|---|---|---|---|
| 4/11 | 4 | $A_1$ | {3549, 3131, 7488, 1315, 4458, 7365, 2855, 2740, 7048, 6229, 8228} | 26931 | 3549+7365+2740+7048+6229 |
| 4/11 | 4 | $A_2$ | {2929, 5455, 4958, 9854, 5140, 6684, 4221, 9644, 3311, 5138, 4929} | 26931 | 2929+5455+4958+5140+3311+5138 |
| 4/11 | 4 | $A_3$ | {2218, 2856, 7934, 5148, 1685, 3161, 2583, 6929, 5654, 3931, 3931} | 26931 | 2218+1685+2583+6929+5654+3931+3931 |
| 4/11 | 4 | $A_4$ | {6597, 1925, 9012, 6079, 9466, 5153, 1897, 3136, 9989, 7530, 1895} | 26931 | 6597+9012+1897+7530+1895 |

**Table 3.** MISSP decryption for C= 7999833427671525772426634417409856716787208454725596462089786782492958755061 6220471110918347425089353477 1926

| n/m | d | sets | Set items | s (plain text) | summations |
|-----|---|------|-----------|----------------|------------|
| 4/9 | 3 | $A_1$ | {799, 983, 342, 767, 152, 577, 242, 663, 441} | 2942 | 342+767+152+577+663+441 |
| 4/9 | 3 | $A_2$ | {740, 985, 671, 678, 720, 845, 472, 559, 646} | 2942 | 985+678+720+559 |
| 4/9 | 3 | $A_3$ | {208, 978, 678, 249, 295, 875, 506, 162, 204} | 2942 | 978+678+249+875+162 |
| 4/9 | 3 | $A_4$ | {711, 109, 183, 474, 250, 893, 534, 771, 926} | 2942 | 711+534+771+926 |

## 5 Conclusion and Future work

This paper introduces the problem of finding a common subset sum for integrated sets (*MISSP*). The objective is to identify a sum that can be achieved by all sets within a 2D set, if such a sum exists. Additionally, a robust cryptosystem utilizing symmetric private keys was proposed for situations where the numbers and sets are of equal sizes. The implementation of this cryptosystem in a communication application was also discussed.

For future work we intend to improve the *MISSP* encryption algorithm using some newer results in solving subset problems such as [20], [21], and [22], and trying to combine the encryption algorithm in other non-standard communication protocols such as TOR (Onion Routing) and solve existing problems as described in [23]. Another interesting aspect of this research is analyzing attack scenarios on this encryption to test its strength. This could help show the robustness of the scheme and demonstrate its strength in juxtaposition to other known encryption schemes.

## Acknowledgements

The authors wish to acknowledge the assistance of Offir Zeevi and Hadar Krispel from Ruppin Academic Center in the implementation of this research, and Ruppin Academic Center for its support in the different stages of this research.

10## References

1. Voloch, N. (2017). MSSP for 2-D sets with unknown parameters and a cryptographic application. Contemp. Eng. Sci., 10(19), 921-931.
2. Garey, M.R., Johnson D.S. (1979)."Computers and Intractability: A Guide to the Theory of NP-Completeness". W.H. Freeman. ISBN 0-7167-1045-5.A3.2: SP13, pg.223.
3. Cormen, T.H.; Leiserson, C. E.; Rivest, R. L.; Stein, Clifford (2001) [1990]. "35.5: The subset-sum problem". Introduction to Algorithms (2nd ed.). MIT Press and McGraw-Hill.
4. Horowitz, E.; Sahni, S. (1974), "Computing partitions with applications to the knapsack problem", Journal of the Association for Computing Machinery, 21: 277–292
5. Pisinger D. (1999). "Linear Time Algorithms for Knapsack Problems with Bounded Weights". Journal of Algorithms, Volume 33, Number 1, October 1999, pp. 1–14
6. Koiliaris, K.; Xu, C. (2015). "A Faster Pseudopolynomial Time Algorithm for Subset Sum". arXiv:1507.02318
7. Johnson, D.S.,( 1974)." Approximation algorithms for combinatorial problems", Journal of Computer and System Sciences, Volume 9, Issue 3, 256-278
8. Kellerer, H. Mansini, R., Pferschy, U., Speranza, M.G. (1997). "An effiicient fully polynomial approximation scheme for the subset-sum problem", Proceedings of the 8th ISAAC Symposium, Springer Lecture Notes in Computer Science 1350, 394-403.
9. Caprara,A. Kellerer,H. , Pferschy,U.(2000) "The Multiple Subset Sum Problem".SIAM Journal on Optimization , Vol. 11, No. 2 : pp. 308-319
10. Woeginger G.J.,Zhongliang Y., (1992)." On the equal-subset-sum problem", Information Processing Letters, Volume 42, Issue 6, Pages 299-302
11. Cieliebak, M., Eidenbenz, S., Pagourtzis, A., & Schlude, K. (2008). "On the Complexity of Variations of Equal Sum Subsets". Nord. J. Comput., 14(3), 151-172.
12. Bazgan, C., Santha, M., Tuza, Z. (1998). "Efficient approximation algorithms for the Subset-Sums Equality problem". In International Colloquium on Automata, Languages, and Programming (pp. 387-396). Springer Berlin Heidelberg.
13. Karimov, M. M., Ochilov, N. N. U., & Tangirov, A. E. (2023). Encryption Methods and Algorithms Based on Domestic Standards in Open-Source Operating Systems. *WSEAS Transactions on Information Science and Applications*, *20*, 42-49.
14. Chahin, N., & Mansour, A. (2023). Improving the IoT and Cloud Computing integration using Hybrid Encryption. *WSEAS Transactions on Design, Construction, Maintenance*, *3*, 1-6.
15. Merkle, R.; Hellman, M. (1978). "Hiding information and signatures in trapdoor knapsacks". Information Theory, IEEE Transactions on. 24 (5): 525–530
16. [14] Martello, S.; Toth, P. (1990). "4 Subset-sum problem". Knapsack problems: Algorithms and computer interpretations. Wiley-Interscience. pp. 105–136.
17. Shamir, A.(1984). "A polynomial-time algorithm for breaking the basic Merkle - Hellman cryptosystem". Information Theory, IEEE Transactions on. 30 (5): 699–704.
18. Kate, A., & Goldberg, I. (2011). "Generalizing cryptosystems based on the subset sum problem". International Journal of Information Security, 10(3), 189-199.
19. https://github.com/nadavvoloch/SUSU_encryption/
20. Antonopoulos, A., Pagourtzis, A., Petsalakis, S., & Vasilakis, M. (2022, March). Faster Algorithms for k-SUBSETSUM and Variations. In Frontiers of Algorithmics: International Joint Conference, IJTCS-FAW 2021, Beijing, China, August 16–19, 2021, Proceedings (pp. 37-52). Cham: Springer International Publishing.
21. Melissinos, N., & Pagourtzis, A. (2018). A faster FPTAS for the subset-sums ratio problem. In Computing and Combinatorics: 24th International Conference, COCOON 2018,